\documentclass[doublecol]{epl2}
\usepackage{amsmath} 
\usepackage{amssymb} 
\usepackage{graphicx}
\usepackage{color}

\title{Multiple State Representation Scheme for Organic
  Bulk Heterojunction Solar Cells: A Novel Analysis Perspective}
\shorttitle{Multiple State Representation Scheme for Organic
  Bulk Heterojunction Solar Cells}
\author{Mario Einax\inst{1,2} \thanks{E-mail: \email{meinax@uos.de}} \and Marcel Dierl\inst{1} \and Philip R. Schiff \inst{2} \and Abraham Nitzan\inst{2}}
\shortauthor{Mario Einax \etal}

\institute{ \inst{1} Fachbereich Physik, Universit\"at Osnabr\"uck,
  Barbarastra{\ss}e 7, 49076 Osnabr\"uck, Germany \\
  \inst{2} School of Chemistry, Tel Aviv University, Tel
  Aviv 69978, Israel}


\abstract{
  The physics of organic bulk heterojunction solar cells is studied
  within a six state model, which is used to analyze the factors that
  affect current-voltage characteristics, power-voltage properties
  and efficiency, and their dependence on nonradiative losses,
  reorganization of the nuclear environment, and environmental polarization.
  Both environmental reorganization and polarity is explicitly taken into
  account by incorporating Marcus heterogeneous and homogeneous
  electron transfer rates. The environmental polarity is found to have a
  non-negligible influence both on the stationary current and on the overall
  solar cell performance. For our organic bulk heterojunction solar
  cell operating under steady-state open circuit condition, we also
  find that the open circuit voltage logarithmically decreases with
  increasing nonradiative electron-hole recombination processes.}

\begin{document}

\maketitle

Considerable progress has been achieved in improving the device
efficiency of bulk heterojunction (BHJ) organic solar cells, with a
recently set record of $10.7\%$. \cite{Service:2012,Heliatek:2012}
Much of this success came about by searching for promising
electron-donor polymers characterized by low optical gap, using
fullerene based electron-acceptor derivatives and optimizing the
interpenetrated frozen-in microstructures. Such phase-separated blend
morphologies are distinguished by a large interface area between the
donor and the acceptor phases, which is a prerequisite to tailor most
efficient organic photovoltaic solar cells (OPVs). While material
design is one successful strategy to improve the OPV setup, another is
to focus on the device physics by developing approaches that take into
account physical and chemical features of BHJ organic solar cells in
order to improve their dynamical operation. In the last two years
there appeared several reviews
\cite{Denner/etal:2009,Deibel/Dyakonov:2010,Nicholson/Castro:2010,Thompson/etal:2011,Nelson:2011}
and perspective articles
\cite{Pensack/etal:2010,Credgington/Durrant:2012} about both material
designs and device physics giving an excellent account of the
state-of-the-art for organic photovoltaics.

In the context of solar energy conversion, device physics aims to
identify routes for improved cell performance by studying models that
account for both the material properties and the underlying
microscopic principles of the energy conversion processes, i.~e.
structural and energetics system parameters, in order to identify
critical factors that affect the overall OPV performance. Thus, the
generated free carriers in a photovoltaic device, which can be
harvested at the electrodes, are limited by the complex interplay
between charge generation, diffusion, and recombination processes. In
BHJ organic solar cells the generation of free charge carriers
requires that photoinduced excitons (bounded electron-hole pairs) on
the donor material must diffuse to and dissociate at the
donor-acceptor (D-A) interface before their recombination takes place.
This exciton dissociation at the D-A interface starts with the
formation of a charge transfer (CT) state (a geminate pair), where the
hole and the electron remain at close proximity on their respective
donor and acceptor sites. \cite{Nelson:2011} This CT state can either
recombine nonradiatively (geminate recombination), or undergo charge
separation leading to mobile electron and hole carriers.
\cite{Nelson:2011,Andersson/etal:2011} However, because of both the
low carrier mobility and the interpenetrated nature of typical BHJ
blends, there is a non-negligible probability that dissociated free
carriers recombine again at the large D-A interface (nongeminate
recombination) before being collected at the electrodes.
\cite{Nelson:2011,Gluecker/etal:2012} These nonradiative
recombinations can be a major loss mechanism that strongly reduces the
power conversion efficiency in BHJ solar cells,
\cite{Nelson/etal:2004,Kirchartz/etal:2009a,Kirchartz/etal:2009b,Giebink/etal:2011,Gruber/etal:2012}
and are mainly influenced by the energy difference between the highest
occupied molecular level $\varepsilon_{\rm D1}$ of D (called HOMO of
D) and the lowest unoccupied level $\varepsilon_{\rm A2}$ of A (called
LUMO of A). Since it was experimentally found for several
donor-acceptor material combinations
\cite{Scharber/etal:2006,Nayak/etal:2011,Nicholson/Castro:2010,Rauh/etal:2011}
that the open circuit voltage $U_{\rm oc}$ is proportional to this
effective energy gap, the nonradiative recombination losses can be
traced back to a drop of $U_{\rm oc}$.

\begin{figure}[h!]
  \centering
\includegraphics[width=0.4\textwidth]{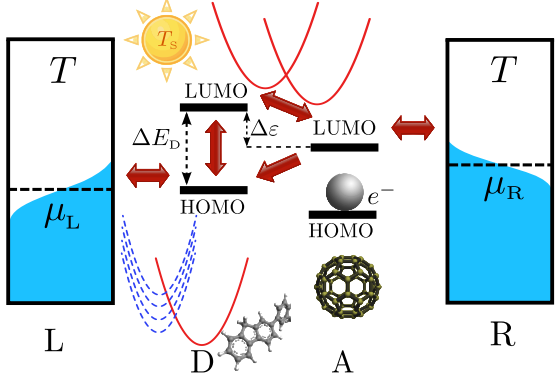}
 \caption{Schematic representation of energetics in BHJ solar cells. The
   system consists of a donor and acceptor, each characterized by
   their HOMO and LUMO levels. The metal-molecule coupling is
   manifested by heterogeneous electron transfer (ET) rates, while
   transitions between the two molecules are controlled by homogeneous
   ET rates, both obtained from the Marcus ET theory.}
 \label{fig:fig1}
\end{figure}

\begin{figure}[h!]
\centering
\includegraphics[width=0.37 \textwidth]{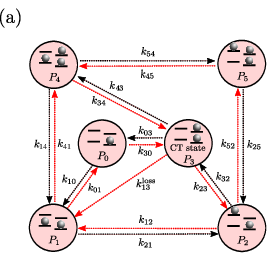} \\

\vspace*{4ex}
\includegraphics[width=0.43 \textwidth]{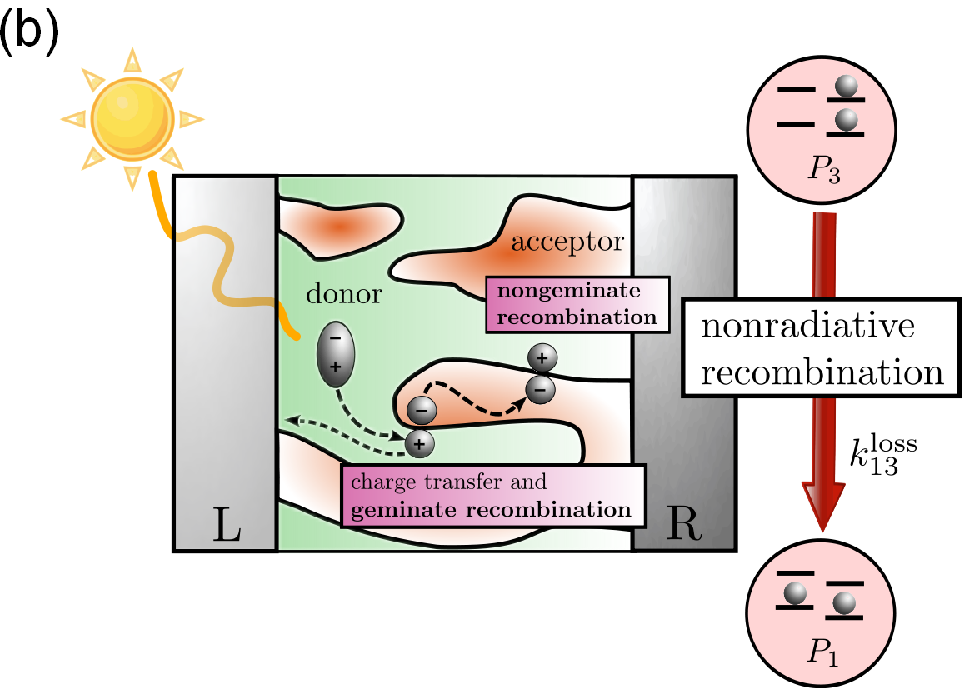}
 \caption{(a) Network representation of
   the underlying master equation associated with the six accessible
   microstates. The graph is composed of six vertices (shown as
   circles). The interconnected vertices represent the probabilities
   $P_j$ to find the system in a microstate $j$ ($j=0,...,5$) and the
   edges connecting some pairs of vertices stand for transitions
   between the states. The edges are drawn as arrows that indicate
   transitions with rate $k_{j'\,j} = k_{j' \leftarrow j}$ from a
   state (vertex) $j$ to $j'$. (b) The recombination process. With a
   certain probability, electrons transferred into the LUMO of the
   acceptor in state $P_3$ can recombine with the hole left on the
   donor (geminate recombination) or migrate in the acceptor phase and
   recombine with the donor at another interfacial position
   (nongeminate recombination). Both processes eventually lead,
   following nonradiative relaxation, to state $P_1$, reducing the
   number of carriers collected at the electrodes.}
 \label{fig:fig2}
\end{figure}

Understanding the relationship between system properties that affect
exciton dissociation at the D-A interface and consequently $U_{\rm
  oc}$ is subject of active research
\cite{Maurano/etal:2010,Rauh/etal:2011,Credgington/Durrant:2012}.
To study how nonradiative losses
at the D-A interface influence the overall device performance, we
invoke the minimal model used in our previous publication
\cite{Einax/etal:2011}, a solar cell with coupled donor and acceptor
molecules, each described as a two-level (HOMO, LUMO) system, in
contact with two electrodes, $\rm L$ and $\rm R$ (see
Fig.~\ref{fig:fig1}). The electrodes are represented by free-electron
reservoirs at chemical potentials $\mu_{K}$ ($K=L, R$) that are set to
$\varepsilon_{\rm\scriptscriptstyle F} =
\varepsilon_{\rm\scriptscriptstyle D1} + \Delta
E_{\rm\scriptscriptstyle D}/2$ in the zero-bias junction. In the model
studied here a symmetric bias is applied with $\mu_{K} =
\varepsilon_{\rm\scriptscriptstyle F} \pm |e| U/2$, where $U$ is the
bias voltage.  Here, we use the notation $\Delta
E_{\rm\scriptscriptstyle K} = \varepsilon_{\rm\scriptscriptstyle K2} -
\varepsilon_{\rm\scriptscriptstyle K1}$ ($K=\rm D,\,A$) for the energy
differences that represent the donor and acceptor band gaps, and refer
to $\Delta\varepsilon = \varepsilon_{\rm\scriptscriptstyle D2} -
\varepsilon_{\rm\scriptscriptstyle A2}$ as the interface or
donor-acceptor LUMO-LUMO gap. It is important to note that these
energies are determined by the detailed electronic structure of the
system. Roughly, $\Delta \varepsilon$ is determined by the single
electron energies augmented by the exciton binding energy - the sum
$E_{\rm\scriptscriptstyle C}=V_{\rm\scriptscriptstyle C} +
V_{\rm\scriptscriptstyle C}'$, where $V_{\rm\scriptscriptstyle C}>0$
is the Coulombic repulsion between two electrons on the acceptor and
$V_{\rm\scriptscriptstyle C}' > 0$ is the Coulombic energy cost to
move an electron away from the donor. Furthermore, interaction with
the nuclear environment, expressed under equilibrium conditions by the
nuclear reorganization energy, affects this energy gap as described
below. The different system states are described by occupation
numbers $n_{\rm\scriptscriptstyle K_j}=0,1$, where $K=D,A$ and
$j=1,2$. In order to use a minimal model that contains essential
physical pictures we limit the number of system states as follows.
First, an imposed restriction $n_{\rm\scriptscriptstyle D1}
n_{\rm\scriptscriptstyle D2} = 0$ ensures that the donor cannot be
double occupied.  In addition we set $n_{\rm\scriptscriptstyle A1}=1$,
so that the acceptor can only receive an additional electron. The
resulting minimal model then consists of six states with respect to
the occupations $(n_{\rm\scriptscriptstyle
  D1},\,n_{\rm\scriptscriptstyle D2},\,n_{\rm\scriptscriptstyle
  A1},\,n_{\rm\scriptscriptstyle A2})$, that we denote by the integers
$j=0,\,...,\,5$, [see Fig.~\ref{fig:fig2}(a)]. Within this six-state
representation, the probability to find the system in state $j$ is
denoted by $P_j$.

This minimal model of a BHJ solar cell accounts for the important
interfacial electronic processes, including excitation, exciton
dissociation, carrier recombination and electron transfer (ET)
processes. In order to focus on these interfacial
processes we disregard in this model exciton diffusion in the donor
and charge carrier diffusion in the acceptor phase, but obviously a
more complete model should take these important processes of the cell
operation into account \cite{Dierl/etal:2014}. In particular, the
ET processes are affected by environmental polarization relaxation.
This is taken here into account
using the nonadiabatic Marcus theory, which is
based on the assumption that in each electronic state the nuclear
motion reaches thermal equilibrium quickly (fast relative to rates of
change of electronic states). The relevant rates are associated with
the following processes:

(a) The electron transfer between the left electrode $\mathrm{L}$ to
the HOMO of the donor is determined by heterogeneous Marcus rates \cite{Nitzan:2006}
[see also Fig.~\ref{fig:fig2}(a)],
\begin{align}
\label{eq:nonhom_MET_A}
k_{10} = k_{43} = \frac{\nu_{\rm\scriptscriptstyle L}}{\sqrt{4 \pi
    \lambda_{\rm\scriptscriptstyle D} k_{\rm B} T}} \int dE\, f(E) e^{
  -\frac{\scriptstyle (E+\mu_{\rm\scriptscriptstyle
      L}-\varepsilon_{{\rm\scriptscriptstyle D1}} -
    \lambda_{\rm\scriptscriptstyle D})^2}{ \scriptstyle 4
    \lambda_{\rm\scriptscriptstyle D} k_{\rm B} T}} \, ,
\end{align}
where $\nu_{\rm\scriptscriptstyle L}$ is the inverse of a
characteristic time scale involved in this process,
$\lambda_{\rm\scriptscriptstyle D}$ is the reorganization energy
associated with the response of the nuclear environment to electronic
population on the donor (also called inner-sphere reorganization energy \cite{Vaissier/etal:2013}),
and $k_{\rm B} T$ is the thermal energy.
Here, $f(x)=1/[\exp(x)+1]$ with $x=E/k_{\rm B} T$ and $T$ is the cell
temperature. A similar expression, with $\mu_{\rm\scriptscriptstyle
  L}$, $\nu_{\rm\scriptscriptstyle L}$, and
$\lambda_{\rm\scriptscriptstyle D}$ replaced by
$\mu_{\rm\scriptscriptstyle R}$, $\nu_{\rm\scriptscriptstyle R}$, and
$\lambda_{\rm\scriptscriptstyle A}$ applies for electron transfer rates
$k_{03}=k_{14}=k_{25}$ from the LUMO of the acceptor to the right
electrode $R$. The corresponding reverse rates, e.g.,
\begin{align}
\label{eq:nonhom_MET_B}
k_{01} = k_{34} = \frac{\nu_{\rm\scriptscriptstyle L}}{\sqrt{4 \pi
    \lambda_{\rm\scriptscriptstyle D} k_{\rm B} T}} \int dE\,
    \tilde{f}(E) e^{ -\frac{\scriptstyle
    (E+\mu_{\rm\scriptscriptstyle
      L}-\varepsilon_{{\rm\scriptscriptstyle D1}} +
    \lambda_{\rm\scriptscriptstyle D})^2}{ \scriptstyle 4
    \lambda_{\rm\scriptscriptstyle D} k_{\rm B} T}} \, ,
\end{align}
where $\tilde{f}(E) = 1- f(E)$, satisfy detailed balance.

(b) Light absorption and molecular excitation take place at the donor
with photon absorption leading to an exciton (electron in an excited
state D2) with rate $k_{21} = k_{54} = \nu_{\rm \scriptscriptstyle{S}}
n_{\rm \scriptscriptstyle{S}} (x_{\rm \scriptscriptstyle{S}})$, where
$n_{\rm{\scriptscriptstyle S}} = 1/\left[\exp(x_{\rm
    {\scriptscriptstyle S}})-1\right]$ with $x_{\rm
  {\scriptscriptstyle S}}=\Delta E_{\rm {\scriptscriptstyle D}}/k_{\rm
  B} T_{\rm {\scriptscriptstyle S}}$ \cite{Rutten/etal:2009}. $T_{\rm {\scriptscriptstyle S}}$
is the sun temperature representing the incident radiation and $\nu_{\rm
  \scriptscriptstyle{S}}$ determines the characteristic time scale for this process.
  Excited states can radiatively decay with rates $k_{12} = k_{45} = \nu_{\rm
  \scriptscriptstyle{S}} ( 1 + n_{\rm \scriptscriptstyle{S}} (x_{\rm
  \scriptscriptstyle{S}}))$.

(c) The dissociation of an exciton at the D-A interface leads to a CT
state $P_3$ with a homogeneous rate given by a Marcus-type expression
\begin{align}
\label{eq:hom_MET_rev}
k_{32} &= \frac{\nu_{\rm\scriptscriptstyle DA}}{\sqrt{4 \pi
    {\lambda}_{{\rm\scriptscriptstyle DA}} k_{\rm B} T}} e^{
  -\frac{\scriptstyle ({\varepsilon}_{\rm\scriptscriptstyle DA} -
    {\lambda}_{{\rm\scriptscriptstyle DA}})^2}{\scriptstyle 4
    {\lambda}_{{\rm\scriptscriptstyle DA}} k_{\rm B} T}} \, , \nonumber\\
k_{23} &= \frac{\nu_{\rm\scriptscriptstyle DA}}{\sqrt{4 \pi
    {\lambda}_{{\rm\scriptscriptstyle DA}} k_{\rm B} T}} e^{
  -\frac{\scriptstyle ({\varepsilon}_{\rm\scriptscriptstyle DA} +
    {\lambda}_{{\rm\scriptscriptstyle DA}})^2}{\scriptstyle 4
    {\lambda}_{{\rm\scriptscriptstyle DA}} k_{\rm B} T}} \, ,
\end{align}
where $\varepsilon_{\rm\scriptscriptstyle DA}=
\varepsilon_{\rm\scriptscriptstyle D2} - (
\varepsilon_{\rm\scriptscriptstyle A2} + E_{\rm\scriptscriptstyle C}
)+ E_{\rm\scriptscriptstyle P}$. $E_{\rm\scriptscriptstyle P}$ is the
term associated with the energy change of charge states, where the
electron is on the donor or on the acceptor. If the environmental
polarization motions (or modes) that respond to charging are different
for the donor and the acceptor species, the reorganization energy for
the ET transfer at the D-A interface is
$\lambda_{\rm\scriptscriptstyle DA}= \lambda_{\rm\scriptscriptstyle
  D}+\lambda_{\rm\scriptscriptstyle A}$ \cite{Nitzan:2006}.
Since the physical origin of $E_{\rm\scriptscriptstyle P}$ and
$\lambda_{\rm\scriptscriptstyle DA}$ is the same, they are related to each other.
The details of this relationship depend on how reorganization occurs on transition
between the initial and final state of the electron transfer process.
In our case, environmental polarization stabilizes the charge separated
state relative to the parent excitonic configuration.
More generally~\cite{note1} if some reorganization exists already in
the absence of environmental polarity it changes according to
$\lambda_{\rm\scriptscriptstyle DA} \rightarrow
\lambda_{\rm\scriptscriptstyle DA} + E_{\rm\scriptscriptstyle P}$.  In
either case the forward rate is not affected by increasing environmental
polarity while the backward process is inhibited.

(d) In the six-state model nonradiative (geminate and nongeminate) recombination of electron
and holes at the D-A interface will together be represented by an effective recombination rate
$k_{13}^{\rm\scriptscriptstyle loss}$ [see Fig.~\ref{fig:fig2}(b)]. However,
future work should take into account the fact that the geminate and
nongeminate rates depend differently on the densities of electrons and
holes, as for example in a rate equation description. In fact, a rate equation description should
imply that the geminate rate is proportional to the number of geminate electron-hole pairs,
while the nongeminate contribution depends on the product of electron
and hole densities at the interface.

The system dynamics is modeled by a master equation approach
accounting for the time evolution of the probabilities $P_j (t)$
($j=0,...,5$) fulfilling normalization $\sum_{j} P_j (t) =1$ at all
times.
\cite{Einax/etal:2010a,Einax/etal:2010b,Dierl/etal:2011,Dierl/etal:2012,Dierl/etal:2013,Sylvester-Hivid/etal:2004}
An elegant network representation \cite{Schnakenberg:1976} can be used
to depict the transitions between the six possible states shown in
Fig.~\ref{fig:fig2}(a). Starting from this, the currents can be
written in terms of state probabilities as follows
\begin{align}
\label{eq:current1}
J_{\rm\scriptscriptstyle L} (t) &= k_{10} ( P_{0} + P_{3})
- k_{01} ( P_{1} + P_{4}) \\
\label{eq:current2}
J_{\rm\scriptscriptstyle R} (t) &= k_{03} ( P_{3} + P_{4} + P_{5} ) - k_{30} ( P_{0} + P_{1} + P_{2} ) \\
\label{eq:current3}
J_{\rm\scriptscriptstyle S}(t) &= k_{21} ( P_{1} + P_{4}) - k_{12} ( P_{2} + P_{5}) \\
\label{eq:current5}
J_{\rm\scriptscriptstyle DA} (t) &= k_{32} P_{2} - k_{23} P_{3}\\
\label{eq:current6}
J_{\rm\scriptscriptstyle loss} (t) \mathbf{}&= k_{13}^{\rm\scriptscriptstyle loss} P_3 \, .
\end{align}
$J_{\rm\scriptscriptstyle L}$ ($J_{\rm\scriptscriptstyle R}$) is the
current entering (leaving) the molecular system from (to) the
electrodes, $J_{\rm\scriptscriptstyle S}$ is the light induced
transition current between the HOMO and the LUMO in the donor phase,
and $J_{\rm\scriptscriptstyle DA}$ is the current between the donor
and acceptor species. $J_{\rm\scriptscriptstyle loss}$ is the loss current
which includes both geminate and nongeminate recombination processes at the D-A
interface. The steady-state solution of the underlying
master equation is given by Kirchhoff's current law
\cite{Kirchhoff:1847} for each state $j=0,..,5$, i.~e. the net influx
must equal the net outflux at each vertex. For example, for $j=0$,
the node condition is $0=k_{01} P_{1}(t) + k_{03} P_{3}(t) -
\left( k_{10}+k_{30}\right)P_{0}(t)$ [see Fig.~\ref{fig:fig2}(a)].
Applying this procedure to each state leads to a closed set of coupled
linear equations. In what follows we specify the steady-state current
$J_{\rm\scriptscriptstyle L} = J_{\rm\scriptscriptstyle R} =
J_{\rm\scriptscriptstyle S} - J_{\rm\scriptscriptstyle loss} =
J_{\rm\scriptscriptstyle DA} - J_{\rm\scriptscriptstyle loss} \equiv
J$.

In order to demonstrate the nature of the kinetics we consider the
following set of parameters:\, $\mu_{\rm\scriptscriptstyle
  L}=\varepsilon_{\rm\scriptscriptstyle D1} + (\Delta
E_{\rm\scriptscriptstyle D} - |e|U)/2$, $\mu_{\rm\scriptscriptstyle
  R}=\mu_{\rm\scriptscriptstyle L} + |e|U$,
$\varepsilon_{\rm\scriptscriptstyle D1} = -0.1\,\rm
eV$,$\varepsilon_{\rm\scriptscriptstyle D2} = 1.4\,\rm eV$,
$\varepsilon_{\rm\scriptscriptstyle A2} = 0.8\,\rm eV$,
$V_{\rm\scriptscriptstyle C} = 0.25\,\rm eV$, and
$V_{\rm\scriptscriptstyle C}' = 0.15\,\rm eV$. Thus, $\Delta
E_{\rm\scriptscriptstyle D} = \varepsilon_{\rm\scriptscriptstyle D2} -
\varepsilon_{\rm\scriptscriptstyle D1} = 1.5\, \rm eV$
\cite{Soci/etal:2007} and $E_{\rm\scriptscriptstyle C}= 0.4\,\rm eV$
\cite{Gregg/etal:2003,Pensack/etal:2010}. For the temperatures we
choose $T=300 \rm K$ and $T_s=6000 \rm K$.  The kinetic rates are set
to $\nu_{\rm\scriptscriptstyle L} = \nu_{\rm\scriptscriptstyle R} =
\nu_{\rm\scriptscriptstyle S} = 0.01 \nu_{\rm\scriptscriptstyle DA}$
and $\nu_{\rm\scriptscriptstyle DA}=10^{12}$s$^{-1}$. For simplicity
we assume that the reorganization energies
$\lambda_{\rm\scriptscriptstyle D}$ and
$\lambda_{\rm\scriptscriptstyle A}$ in the donor and acceptor phase
are both $0.1\,\rm eV$.
\begin{figure}[h!]
\centering
\includegraphics[width=0.45 \textwidth]{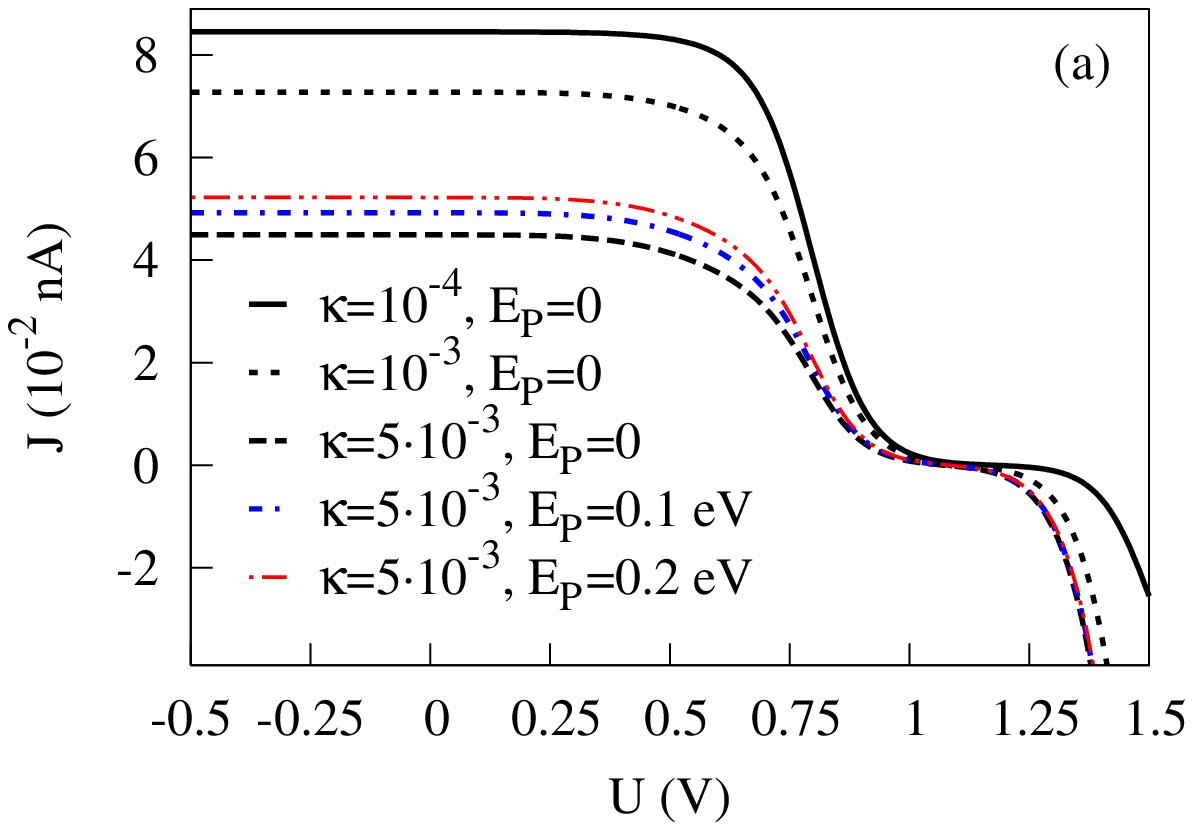}\\
\vspace*{2ex}
\includegraphics[width=0.45 \textwidth]{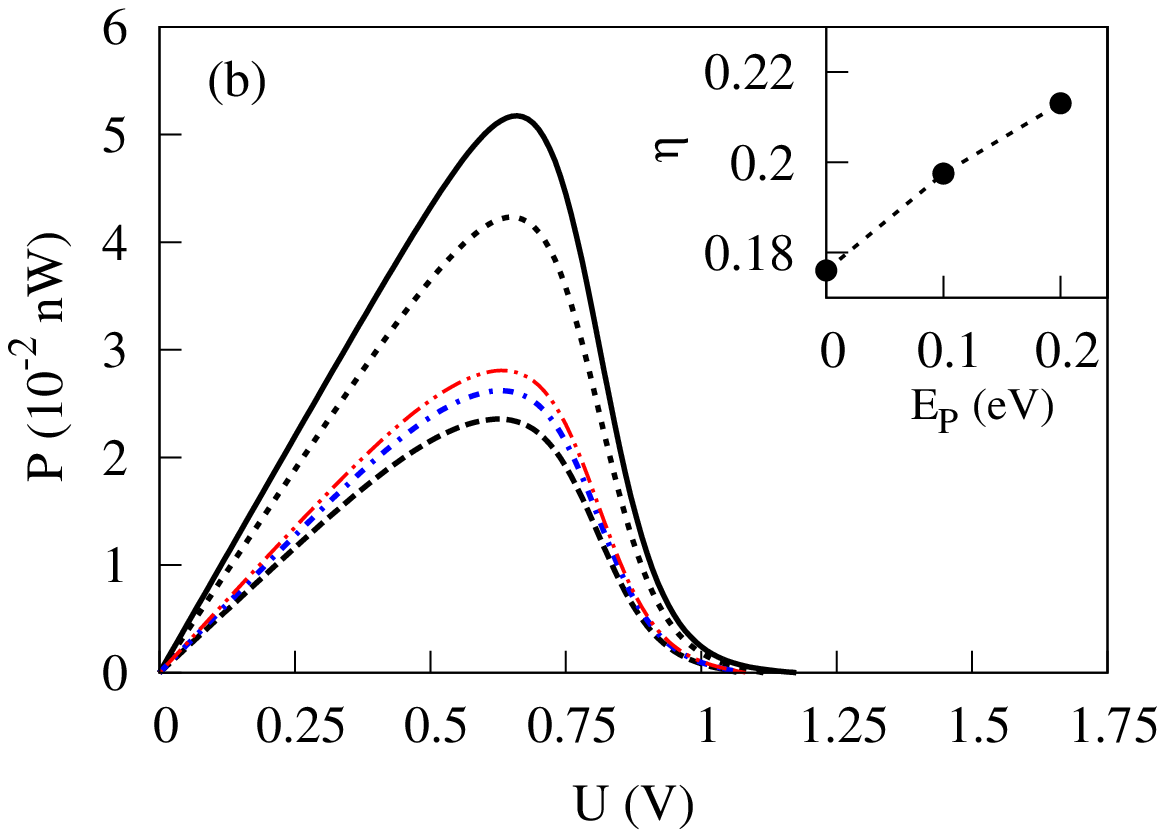}
 \caption{(a) Steady-state current and
   (b) power as functions of bias voltage for three different rate
   ratios $\kappa = k_{13}^{\rm\scriptscriptstyle loss}/k_{32}$ and
   polarity energies $E_{\rm\scriptscriptstyle P}$. In (b) the assignment of lines
   is given in the legend (a). The inset in (b) shows the thermodynamic
   efficiency at maximum power \cite{Einax/etal:2011} as function of $E_{\rm\scriptscriptstyle P}$ for $\kappa = 5\cdot 10^{-3}$.}
 \label{fig:fig3}
\end{figure}

The behaviors of the current- and power-voltage characteristics as
functions of both the nonradiative losses and environmental
polarity are shown in Fig.~\ref{fig:fig3}. Figure~\ref{fig:fig3}(a) shows the
current-voltage characteristics for different
$E_{\rm\scriptscriptstyle P}$ values and several rate ratios $\kappa =
k_{13}^{\rm\scriptscriptstyle loss}/k_{32}$, which measure the
magnitude of the recombination losses compared to the fast ET transfer
at the D-A interface. Obviously the steady-state current decreases
with increasing losses. In particular, the plateau in the $J(U)$ curve
drops down with larger $k_{13}^{\rm\scriptscriptstyle loss}$.
For a certain applied voltage value $U_{\rm oc}$ the current is zero
and becomes negative for $U>U_{\rm oc}$. To demonstrate the influence
of the environmental polarity we choose $\kappa = 5\cdot 10^{-3}$ and two
different polarity energies $E_{\rm\scriptscriptstyle P}=0.1$\,eV and
$0.2$\,eV. Compared to the curve corresponding to
$E_{\rm\scriptscriptstyle P}=0$ and $\kappa = 5\cdot 10^{-3}$ we find
a significant increase in the current. Figure~\ref{fig:fig3}(b)
compares the performance $P(U)=U J(U)$ of the BHJ solar cell for
various loss currents and environmental polarity energies.
Clearly, the curves in Fig.~\ref{fig:fig3}(b) exhibit a maximum in the region where
the current in Fig.~\ref{fig:fig3}(a) decreases sharply. We also
observe that the performance of the solar cell strongly depends on
both environmental polarization relaxation and nonradiative losses. In particular, the
inset in Fig.~\ref{fig:fig3}(b) shows the dependence of the
thermodynamic efficiency $\eta$ on $E_{\rm\scriptscriptstyle P}$ for $\kappa = 5\cdot 10^{-3}$.
Because the maximum power output is shifted to larger $U$ by increasing $E_{\rm P}$,
the solar cell becomes more efficient
by increasing the environmental polarization.

Next, we analyze the steady-state performance of the BHJ solar cell
operating under open circuit condition [$J(U_{\rm oc})=0$].
\begin{figure}[h!]
\centering
\includegraphics[width=0.45 \textwidth]{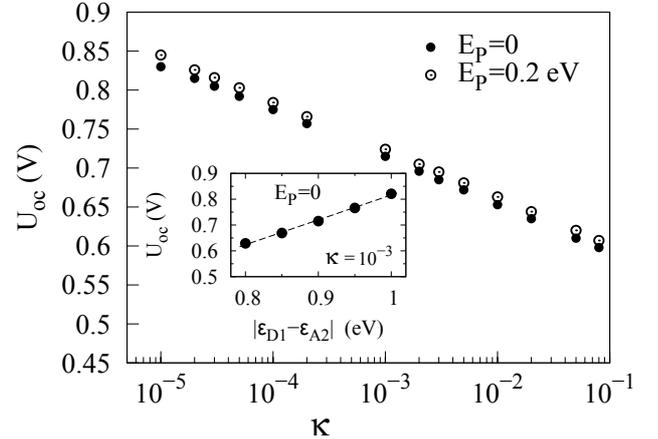}
\caption{Open circuit voltage as function of rate
  ratio $\kappa$ for vanishing $E_{\rm\scriptscriptstyle P}$ (filled circles)
  and for $E_{\rm\scriptscriptstyle P}=0.2$\,eV (open circles).  The inset
  shows $U_{\rm oc}$ as function of the effective band gap
  $|\varepsilon_{\rm\scriptscriptstyle D1}-\varepsilon_{\rm\scriptscriptstyle A2}|$ of the D-A blend for $\kappa=10^{-3}$.}
 \label{fig:fig4}
\end{figure}
Figure~\ref{fig:fig4} shows $U_{\rm oc}$ calculated for the chosen
parameter set as function of $\kappa$.  Increasing nonradiative loss
rates leads to a decrease in the open circuit voltage,
in agreement with earlier findings \cite{Maurano/etal:2010,Deibel/Dyakonov:2010}
As a result, we find that $U_{\rm oc}$ is proportional to $\ln(\kappa)$ [see filled circles for
$E_{\rm\scriptscriptstyle P}=0$ in Fig.~\ref{fig:fig4}]. It is also
interesting to see how these results change upon increasing environmental
polarity, in particular because it inhibits recombination but at the
same time makes a larger LUMO-LUMO gap. The open circuit voltage
increases by considering environmental polarity [see Fig.~\ref{fig:fig4}].
In the inset of Fig.~\ref{fig:fig4} we study the
influence of the effective band gap
$|\varepsilon_{\rm\scriptscriptstyle
  D1}-\varepsilon_{\rm\scriptscriptstyle A2}|$ on $U_{\rm oc}$.
We observe that our BHJ solar cell model well accounts for the present common
understanding that $U_{\rm oc}$ is linearly proportional to the
effective bandgap. For example, let us consider a donor HOMO level of
$\varepsilon_{\rm\scriptscriptstyle D1}=-0.1$ \,eV and a rate ratio
$\kappa=k_{13}^{\rm\scriptscriptstyle loss}/k_{32}=10^{-3}$, where we
vary $k_{32}$ and consequently also the acceptor LUMO level
$\varepsilon_{\rm\scriptscriptstyle A2}$. We then have $U_{\rm oc} = |
\varepsilon_{\rm\scriptscriptstyle D1}-\varepsilon_{\rm\scriptscriptstyle A2}|/|e| - a(\kappa)$. The value of the
parameter $a$ depends on $\kappa$ \cite{Koster/etal:2012}. For $\kappa=10^{-3}$, the fit yields $a(\kappa=10^{-3})=0.17$,
which corresponds to the dashed line in the inset of Fig.~\ref{fig:fig4}.

In conclusion, we developed a model using a six-state representation
for a D-A blend which accounts for essential aspects of the device
physics of BHJ solar cells. In particular, we incorporated more
realistically the molecule-reservoir coupling in terms of
heterogeneous Marcus rates and the ET process at the D-A interface by
homogeneous Marcus rates. In addition we have demonstrated how
nonradiative (geminate and nongeminate) processes can be captured
within a six-state representation. We also discussed the influence of
environmental polarity on the charge separation process. The
environmental polarity turned out to have a decisive influence on both the
stationary current and the overall solar cell performance.
Focusing on a BHJ cell operating under steady-state open circuit condition, we
observed a less pronounced influence of the environmental polarity on $U_{\rm oc}$ and
found $U_{\rm oc}$ proportional to $\ln(\kappa)$ by varying the rate ratio $\kappa$. Regarding future
generalizations that include hot electrons
\cite{Pensack/etal:2010,Grancini/etal:2013,Jailaubekov/etal:2013} and
tail states \cite{Nayak/etal:2011}, the present model provides a
framework for analyzing the open circuit voltage behavior.

\section{Acknowledgement}
  We thank Carsten Deibel for illuminating discussions on geminate and
  nongeminate recombination processes, and an anonymous referee for useful comments.
  M.E. gratefully acknowledges funding by a Minerva Short-Term Research Grant. The research
  of A.N. is supported by the Israel Science Foundation, the Israel-US Binational Science
  Foundation (grant No. 2011509), and the European Science Council (FP7/ERC grant No. 226628).

\bibliographystyle{eplbib}
\bibliography{OPV_cells}

\begin{thebibliography}{10}
\bibitem{note1}
On transition from state a $1$ to state a $2$, if the two corresponding
  parabolic nuclear potential surfaces are $V_1(x) = E_1 + (1/2)k x^2$ and
  $V_2(x) = E_2 + (1/2)k (x-\lambda)^2$, introducing an additional
  environmental polarity response such that $V_2^{\rm new}(x)=V_2(x) - px$
  implies $V_2^{\rm new}(x) = E_2 -p^2/2k + (1/2)k (x-\lambda-p/k)^2$ so that
  $E_1-E_2 \rightarrow E_1-E_2+p^2/2k$ and $\lambda \rightarrow \lambda +p^2/2k
  = \lambda + E_P$.
\end{thebibliography}

\end{document}